\documentclass[a4paper,11pt]{article}
\usepackage{pos}

\title{Critical point in heavy-quark region of QCD on fine lattices}

\author*[a,b]{Masakiyo Kitazawa}
\author[c]{Ryo Ashikawa}
\author[d]{Shinji Ejiri}
\author[e]{Kazuyuki Kanaya}
\author[f]{Hiroto Sugawara}

\affiliation[a]{Yukawa Institute for Theoretical Physics, \\
  Kyoto University, Kyoto 606-8502, Japan}

\affiliation[b]{J-PARC Branch, KEK Theory Center, 
  Institute of Particle and Nuclear Studies, KEK, \\
  Tokai, Ibaraki 319-1106, Japan}

\affiliation[c]{Department of Physics, Osaka University, \\
  Toyonaka, Osaka 560-0043, Japan}

\affiliation[d]{
  Department of Physics, Niigata University,\\
  Niigata 950-2181, Japan}

\affiliation[e]{
  Tomonaga Center for the History of the Universe, University of Tsukuba,\\
  Tsukuba, Ibaraki 305-8571, Japan}

\affiliation[f]{
Graduate School of Science and Technology, Niigata University, \\
Niigata 950-2181, Japan
}

\emailAdd{kitazawa@yukawa.kyoto-u.ac.jp}

\abstract{
  We perform a finite-size scaling analysis of the critical point in the heavy-quark region of QCD at nonzero temperature. Our previous analysis on the Binder cumulant at $N_t=4$ is extended to finer lattices with $N_t=6$ and $8$. The aspect ratio is also extended up to $15$ to suppress the non-singular contribution. High-precision analysis of the Binder cumulant is realized by an efficient Monte-Carlo simulation with the hopping-parameter expansion (HPE). Effects of higher-order terms in the HPE are incorporated by the reweighting method.
}

\FullConference{The 40th International Symposium on Lattice Field Theory (Lattice 2023)\\
July 31st - August 4th, 2023\\
Fermi National Accelerator Laboratory\\}

\begin{document}
\maketitle

\section{Introduction}

An interesting feature of the medium described by Quantum Chromodynamics (QCD) is that appearances of critical points (CPs) are expected with variations of various parameters such as temperature ($T$), baryon chemical potential ($\mu_{\rm B}$), and quark masses. CPs are expected to exist on the QCD phase diagram on the $T$--$\mu_{\rm B}$ plane with physical quark masses. Their experimental search is one of the main goals of relativistic heavy-ion collisions~\cite{Asakawa:2015ybt,Bluhm:2020mpc}. CPs are also expected to appear when the quark masses are varied from the physical ones at $\mu_{\rm B}=0$ both in the light- and heavy-quark regions~\cite{Philipsen:2021qji}. However, their locations and even existence remain controversial~\cite{Ejiri:2019csa,Shirogane:2020muc,Cuteri:2020yke,Kuramashi:2020meg,Dini:2022,Kiyohara:2021smr,Wakabayashi:2021eye}.

In this proceeding, we report on our recent study of the (fourth-order) Binder-cumulant analysis of the CP in the heavy-quark region~\cite{Kiyohara:2021smr,prep}. The Binder cumulant, $B_4$, is a useful quantity to investigate the CPs in numerical simulations~\cite{Binder:1981sa}. It is known from the finite-size scaling argument that $B_4$ obtained on numerical simulations with different spatial volumes has a crossing at the CP. Moreover, the behavior of $B_4$ around the crossing point is determined by the universality class to which the CP belongs. One thus can investigate the universality class of a CP with the use of $B_4$.

In our previous study Ref.~\cite{Kiyohara:2021smr}, we investigated the behavior of $B_4$ near the CP in the heavy-quark region at the temporal lattice extent $N_t=4$. To accelerate the numerical simulations, we employed the simulation based on the hopping-parameter expansion (HPE). The Monte-Carlo simulations are performed with respect to the action at leading order (LO) of the HPE, and the effects at next-to-leading (NLO) order are incorporated in the analysis by the reweighting method. We found that this method works quite effectively to obtain high-precision data up to the aspect ratio $N_s/N_t=LT=12$, where $L=N_s a$ is the spatial lattice extent in physical units and $T=1/N_t a$ is the temperature with $a$ the lattice spacing.

In this proceeding, we extend the analysis to $N_t=6$ and $8$~\cite{prep}. We find that the violation of the finite-size scaling of $B_4$ becomes stronger as $N_t$ becomes larger, which suggests that larger aspect ratios are needed to investigate the CP on finer lattices properly.

\section{Formalism}

We start from the lattice action $S = S_{\rm g} + S_{\rm q}$ where
the Wilson gauge action $S_{\rm g}$ and Wilson quark action $S_{\rm q}$ are
\begin{align}
  S_{\rm g} = -6 N_{\rm site} \,\beta \, \hat{P},
  \qquad
  S_{\rm q} = N_{\rm f} \sum_{x,\,y} \bar{\psi}_x \,
  M_{xy} (\kappa) \, \psi_y ,
  \label{eq:S}
\end{align} 
for degenerate $N_{\rm f}$ flavors of quarks, 
$\hat{P}$ is the plaquette operator, $N_{\rm site} = N_s^3 \times N_t$ is the lattice space-time volume, $\beta = 6/g^2$ is the gauge coupling, $\kappa$ is the quark hopping parameter, and 
\begin{align}
  M_{xy} (\kappa)
  =&\ \delta_{xy} - \kappa B_{xy},
  \label{eq:Mxy}
  \\
  B_{xy}
  =&\  \sum_{\mu=1}^4 \left[ (1-\gamma_{\mu})\,U_{x,\mu}\,\delta_{y,x+\hat{\mu}} + (1+\gamma_{\mu})\,U_{y,\mu}^{\dagger}\,\delta_{y,x-\hat{\mu}} \right],
  \label{eq:B}
\end{align} 
is the Wilson quark kernel. In the following, we consider the case $N_{\rm f}=2$. Generalization to other $N_{\rm f}$ or non-degenerate cases is easy with our method~\cite{Kiyohara:2021smr}.

In the heavy-quark region $\kappa\ll1$, it is a good approximation to express $\ln{\rm det}M$ by the hopping-paramter expansion (HPE) as 
$\ln \det M(\kappa)  =-\sum_{n=1}^{\infty} (1/n) {\rm Tr} 
  \left[ B^n \right] \kappa^n$.
The $n$th-order term in the HPE is graphically represented by the closed trajectories of length $n$~\cite{Kiyohara:2021smr}, and one obtains the effective action
\begin{align}
  S_{\rm eff}
  =&\ S_{\rm g} -N_{\rm f} \ln \det M = S_{\rm g} - N_{\rm f} N_{\rm site} \sum_{n=1}^\infty \big( \hat W(n) + \hat L(N_t,n) \big) \kappa^n
  \\
  =&\ S_g + S_{\rm LO} + S_{\rm NLO} + \cdots,
    \label{eq:SeffWL}
\end{align}
where $\hat W(n)$ and $\hat L(N_t,n)$ are contributions from trajectories without and with windings along the temporal direction, provided that the spatial extent is sufficiently large. Here, $S_{\rm LO}$ and $S_{\rm NLO}$ are the leading-order (LO) and next-to-leading order (NLO) contributions defined by
\begin{align}
  S_{\rm LO} =&\ - N_{\rm f} N_{\rm site} \big( \hat W(4) \kappa^4 + \hat L(N_t,N_t) \kappa^{N_t} \big),
  \label{eq:LO}
  \\
  S_{\rm NLO} =&\ - N_{\rm f} N_{\rm site} \big( \hat W(6) \kappa^6 + \hat L(N_t,N_t+2) \kappa^{N_t+2}\big),
\end{align}
where $\hat W(4) = 288 \hat P$ and 
$\hat L(N_t,N_t) = (2^{N_t+1} N_{\rm c}/N_t) {\rm Re}\hat\Omega$ with the Polyakov loop $\hat\Omega$. 
In the $S_{\rm NLO}$, $\hat W(6)$ is given by the 6-step Wilson loops, and $\hat L(N_t,N_t+2)$ is given by $(N_t+2)$-step bent Polyakov loops~\cite{Ejiri:2019csa}.

\section{Numerical setup}

We perform Monte-Carlo simulations for the leading-order action
\begin{align}
S_{g+\rm LO}=S_{\rm g}+S_{\rm LO}
  = -6 N_{\rm site} \beta^* \hat{P} - \lambda N_s^3 {\rm Re}\hat\Omega,
\end{align}
with $\beta^* = \beta + 48 N_{\rm f} \kappa^4$ 
and
$\lambda = 2^{N_t+1} N_{\rm c} N_{\rm f} \kappa^{N_t}$, 
and incorporate the effects of $S_{\rm NLO}$ by the reweighting.
Gauge configurations are generated by the pseudo heat bath and over relaxation algorithms. We found that this method enables high-precision numerical analyses around the CP in the heavy-quark region thanks to efficient Monte-Carlo updates avoiding the overlapping problem in reweighting~\cite{Kiyohara:2021smr}. 

The temporal and spatial lattice sizes are varied within the range $N_t=4,6,8$ and $N_s/N_t=LT\le15$.
For each $(N_s,N_t)$, gauge configurations are generated
for $3$--$6$ sets of $(\beta^*,\lambda)$, which are chosen to be close to the transition point at LO. For each parameter, we perform the measurements typically $10^6$ times.
The statistical errors of observables are estimated by the jackknife method with the binsize of $10,000$--$25,000$ measurements, which is sufficiently larger than the estimated autocorrelation lengths.

\section{Numerical results}

In this study, to investigate the CP in the heavy-quark region we study the Binder cumulant of the real part of the Polyakov loop $\hat\Omega_{\rm R}={\rm Re}\hat\Omega$,
\begin{align}
    B_4 = \frac{\langle \hat\Omega_{\rm R}^4 \rangle_{\rm c}}{\langle \hat\Omega_{\rm R}^2 \rangle_{\rm c}^2} +3
    = \frac{\langle\delta\hat\Omega_{\rm R}^4 \rangle}{\langle\delta\hat\Omega_{\rm R}^2 \rangle^2},
    \label{eq:B4}
\end{align}
on the transition line as a function of $\lambda$.
It is known that this quantity observed at different spatial volumes has a crossing at the CP~\cite{Binder:1981sa}. Moreover, from the finite-size scaling it is known that this quantity behaves around the CP of the $Z(2)$ universality class, to which the QCD critical point is believed to belong, as
\begin{align}
  B_4(\lambda,LT) =&\ b_4 + c ( \lambda-\lambda_{\rm c} ) (LT)^{1/\nu} ,
  \label{eq:B4fit4}
  \\
  b_4=&\ 1.604, \mbox{~and~} \nu=0.630.
  \label{eq:b4nu}
\end{align}

\begin{figure}[t]
 \centering
 \includegraphics[width=0.48\textwidth, clip]{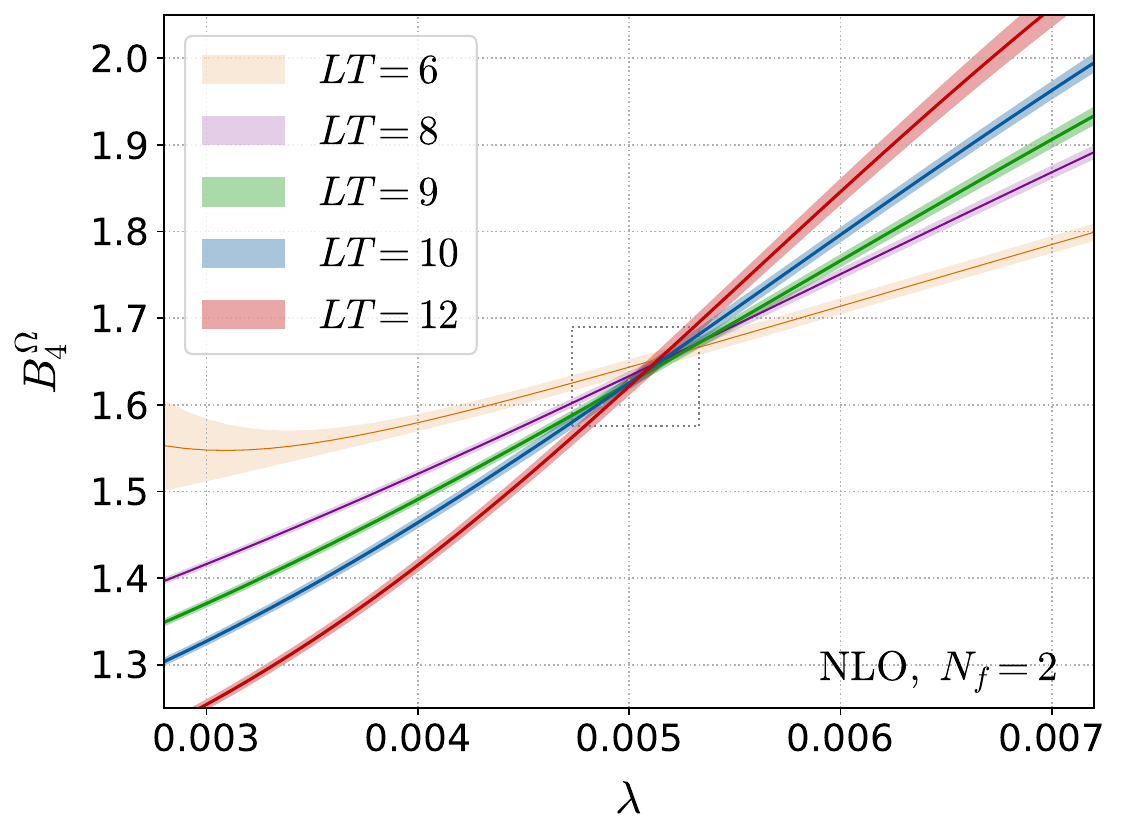}
 \includegraphics[width=0.48\textwidth, clip]{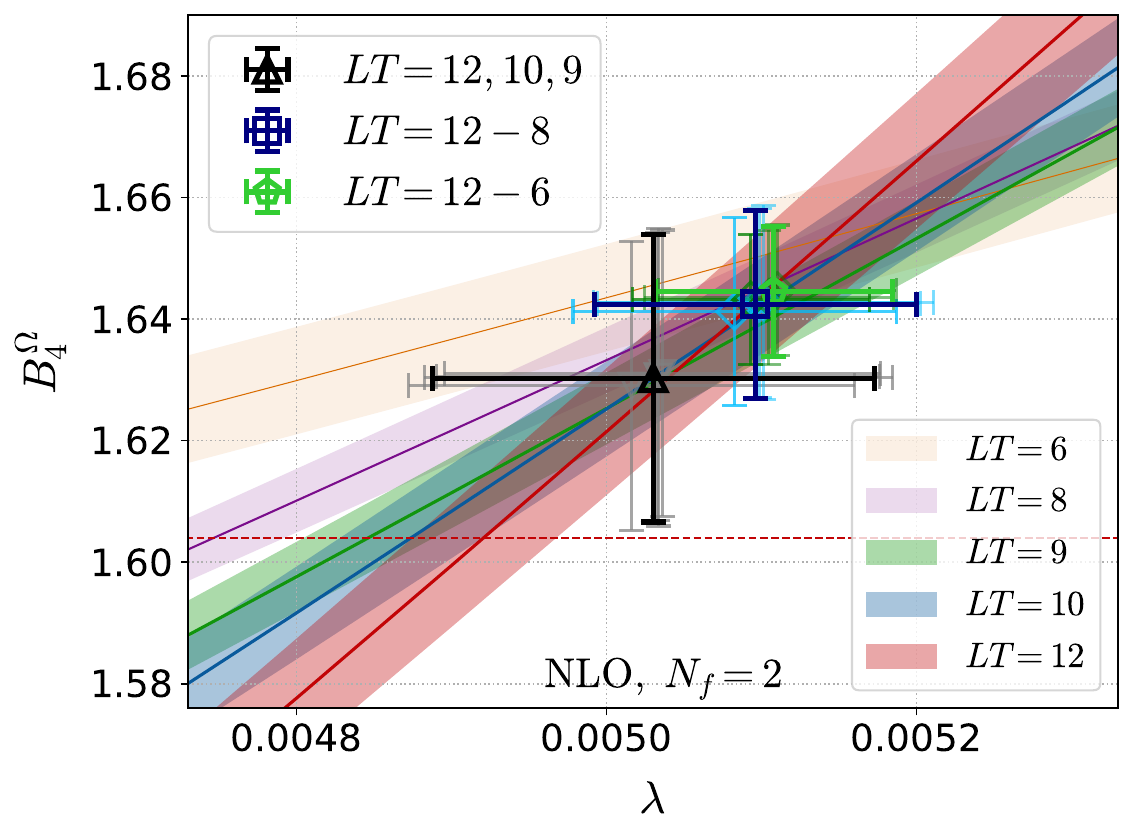}
 \caption{
   $\lambda$ dependence of the Binder cumulant near the CP in the heavy-quark region at $N_t=4$~\cite{Kiyohara:2021smr}.
   The right panel is an enlargement of the left one near the crossing point.
 }
 \label{fig:4}
\end{figure}

In Fig.~\ref{fig:4}, we first show our previous result at $N_t=4$ on 
the Binder cumulant $B_4^\Omega=B_4$ along the transition line as a function of $\lambda$
for five values of $LT$~\cite{Kiyohara:2021smr}.
The right panel is an enlargement of the left panel around the crossing point.
The figure shows that $B_4$ has a crossing at
$\lambda=\lambda_{\rm c}\simeq0.005$, suggesting the existence of the CP around there.
The result for $LT=6$ has a small but visible deviation from the crossing, while the larger volume results at $LT\ge8$ are more stable.

To determine the parameters in Eq.~\eqref{eq:B4fit4} quantitatively, 
we fit the numerical results of $B_4$ by the fitting function~\eqref{eq:B4fit4} 
with $b_4$, $\lambda_{\rm c}$, $\nu$, $c$ being the fit parameters. 
The result of the fit gives~\cite{Kiyohara:2021smr}
\begin{align}
  &b_4 = 1.630(24)(2), \quad \nu=0.614(48)(3), \quad \lambda_c=0.00503(14)(2),
  \quad {\rm for~} LT=12,10,9,
  \\
  &b_4 = 1.643(15)(2), \quad \nu=0.614(29)(3), \quad \lambda_c=0.00510(10)(2),
  \quad {\rm for~} LT=12,10,9,8,
\end{align}
which are indicated by open symbols with error bars in the right panel of Fig.~\ref{fig:4}.
One finds that the values of $b_4$ and $\nu$ obtained with the three largest volumes, $LT=12,10,9$, are consistent with those in the $Z(2)$ universality class, Eq.~\eqref{eq:b4nu}, almost within statistics. However, the result including $LT=8$ has statistically-significant deviations from the $Z(2)$ values. This result suggests that the violation of the finite-size scaling is not suppressed well even at $LT=8$.

\begin{figure}[t]
 \centering
 \includegraphics[width=0.48\textwidth, clip]{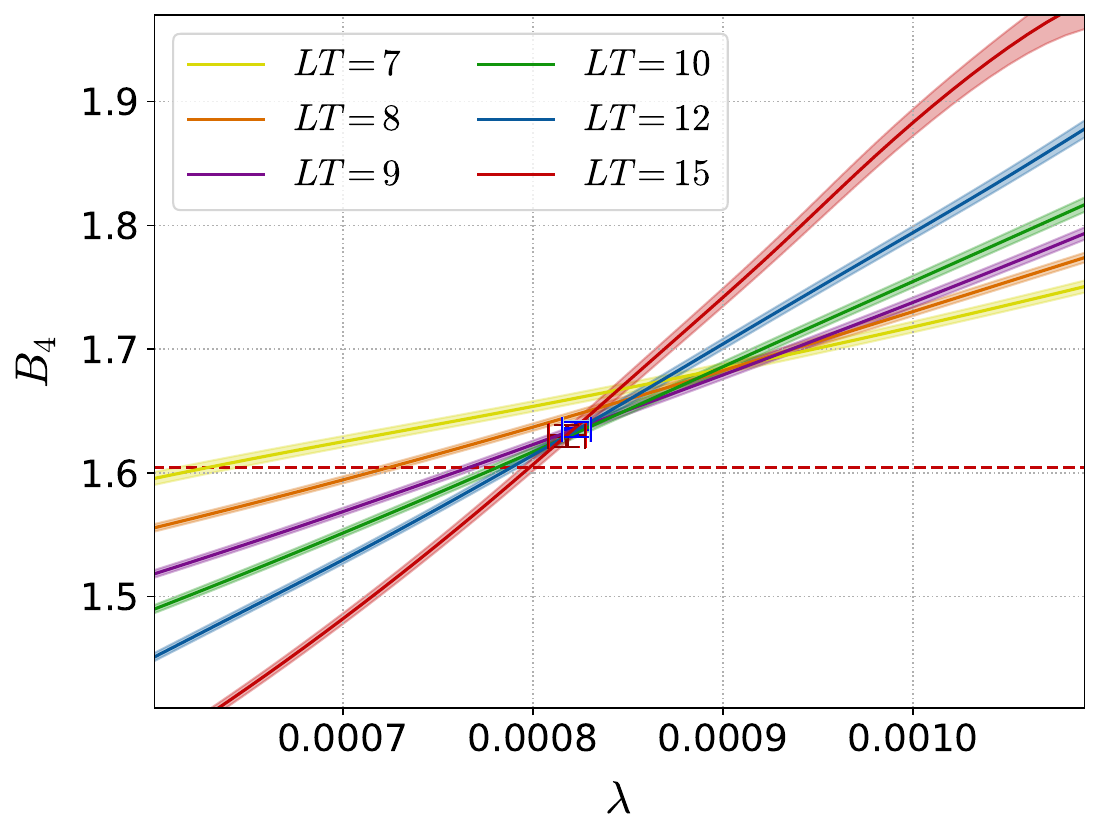}
 \includegraphics[width=0.48\textwidth, clip]{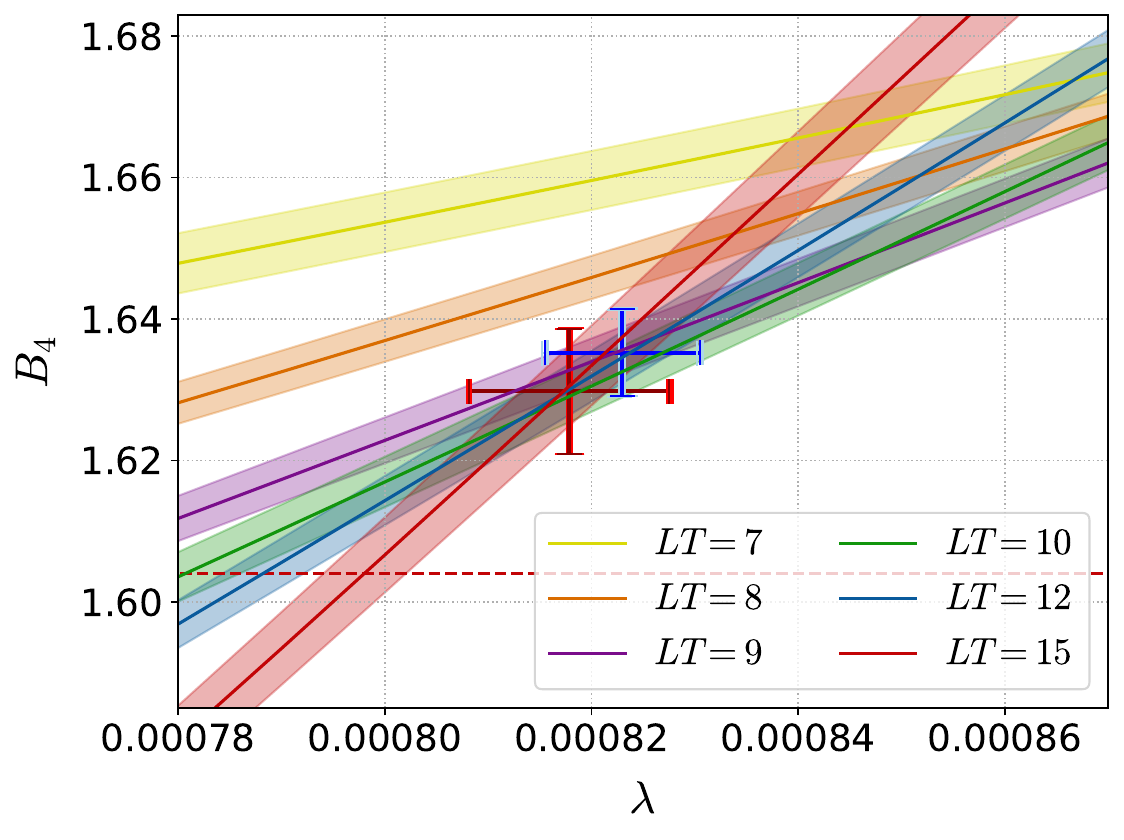}
 \caption{
   $\lambda$ dependence of the Binder cumulant near the CP in the heavy-quark region at $N_t=6$~\cite{prep}.
   The right panel is an enlargement of the left one near the crossing point.
 }
 \label{fig:6}
\end{figure}

Next, let us see our new results at $N_t=6$~\cite{prep}.
In Fig.~\ref{fig:6}, we show the $\lambda$ dependence of $B_4$ obtained at the several aspect ratios up to $LT=15$ at $N_t=6$.
The right panel is again the enlargement of the left one around the crossing point.
From the figure, one sees that the crossing of $B_4$ occurs around $\lambda\simeq0.00082$, but the result at $LT=8$ has a statistically significant deviation from the crossing point of the results of yet larger $LT$. This result indicates that the violation of the finite-size scaling at the same $LT$ becomes larger as the lattice spacing becomes finer.

To determine the parameters in Eq.~\eqref{eq:B4fit4}, we performed the four-parameter fit to the numerical results in Fig.~\ref{fig:6} with Eq.~\eqref{eq:B4fit4}.
The result of the fit with the three largest volumes $LT=15,12,10$ is 
\begin{align}
  &b_4 = 1.630(9), \quad \nu=0.624(19), \quad \lambda_c=0.000818(10),
  \qquad {\rm for~} LT=15,12,10.
  \label{eq:B4Nt6}
\end{align}
One sees that the value of $b_4$ has a statistically-significant deviation from Eq.~\eqref{eq:b4nu}, while the value of $\nu$ is consistent with the $Z(2)$ universality class. This result may imply that $LT=10$ is not large enough to suppress the violation of the finite-size scaling at $N_t=6$.

By converting the value of $\lambda_c$ in Eq.~\eqref{eq:B4Nt6} to the hopping parameter, we obtain $\kappa_c^{\rm NLO}=0.09003(19)$. Using the method proposed in Ref.~\cite{Wakabayashi:2021eye} to incorporate the yet higher order contributions of the HPE, this value is converted to $\kappa_c=0.08781(17)$~\cite{prep}.
In Ref.~\cite{Cuteri:2020yke}, the value of $\kappa_c$ has been investigated by the Monte Carlo simulations with dynamical fermions and they obtained $\kappa_c=0.0877(9)$. Our result is consistent with this previous result within statistics.

\begin{figure}[t]
 \centering
 \includegraphics[width=0.325\textwidth, clip]{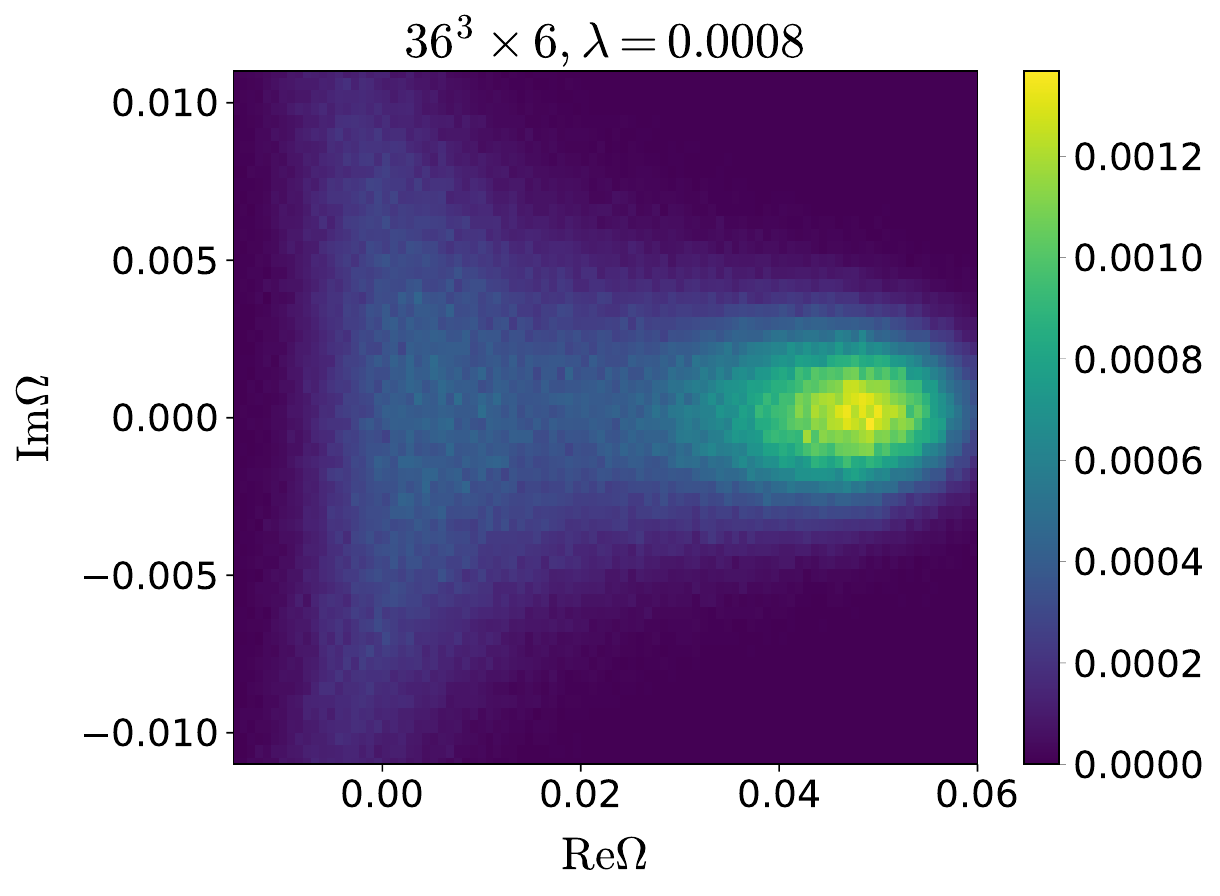}
 \includegraphics[width=0.325\textwidth, clip]{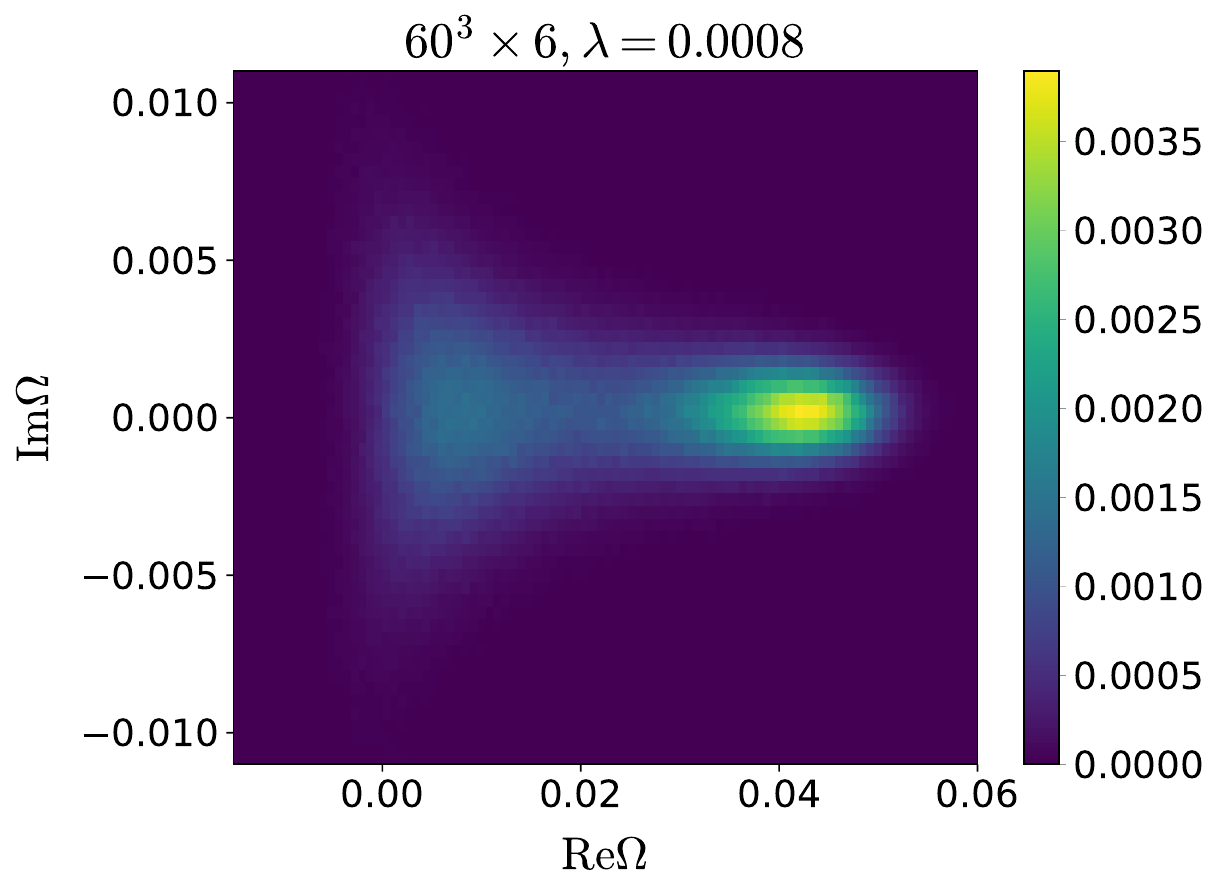}
 \includegraphics[width=0.325\textwidth, clip]{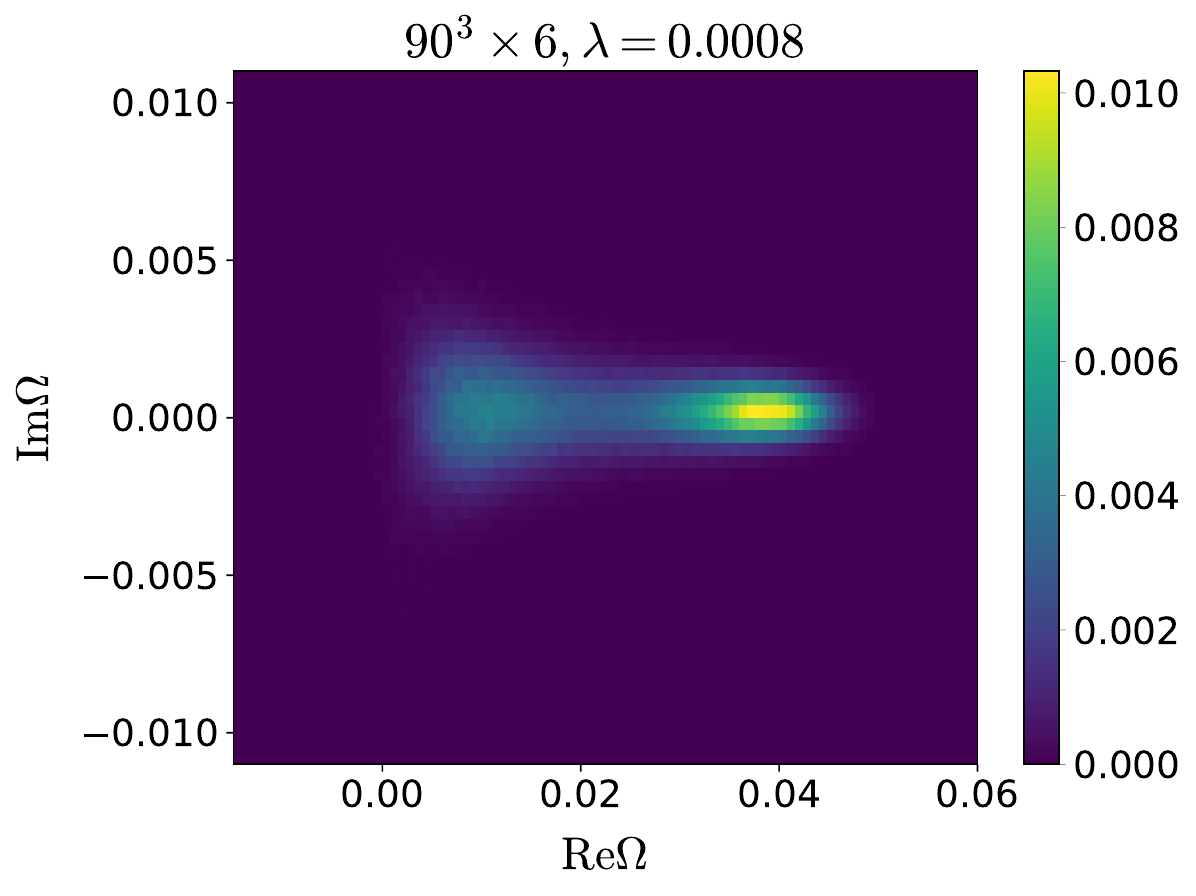}
 \caption{
   Distribution of the Polyakov loop $\hat\Omega$ at the CP at $N_t=6$~\cite{prep}.
 }
 \label{fig:Omega}
\end{figure}

To understand the origin of the strong violation of the finite-size scaling at $N_t=6$, in Fig.~\ref{fig:Omega} we show the contour map of the distribution of $\hat\Omega$ on the complex plane near the CP at $LT=6$ (left), $10$ (middle), $15$ (right). From the left panel, one finds that the distribution has a triangular form at $LT=6$, which is understood as the remnant of $Z(3)$ center symmetry at $\lambda=0$. As a result, the distribution of $\hat\Omega_{\rm R}$ is not symmetric between two peaks. One thus can understand that this causes the violation of the finite-size scaling of the $Z(2)$ universality class. The remnant of $Z(3)$ symmetry is visible even at $LT=10$ and $15$, although the distribution approaches a symmetric one as $LT$ becomes larger.
One possible explanation for why such an effect is more prominent at $N_t=6$ than $N_t=4$ is to attribute the result to the strength of the first-order phase transition at $\lambda=0$. It is known that the latent heat in pure $SU(3)$ YM theory is large at $N_t=4$ due to the lattice artifact~\cite{Shirogane:2020muc}. This fact implies the stronger first-order transition at $\lambda=0$ and that a stronger external field is needed to make the transition crossover. As a result, the CP is located at larger $\lambda$ at $N_t=4$, where the effects of the original $Z(3)$ symmetry are more suppressed. 

\begin{figure}[t]
 \centering
 \includegraphics[width=0.48\textwidth, clip]{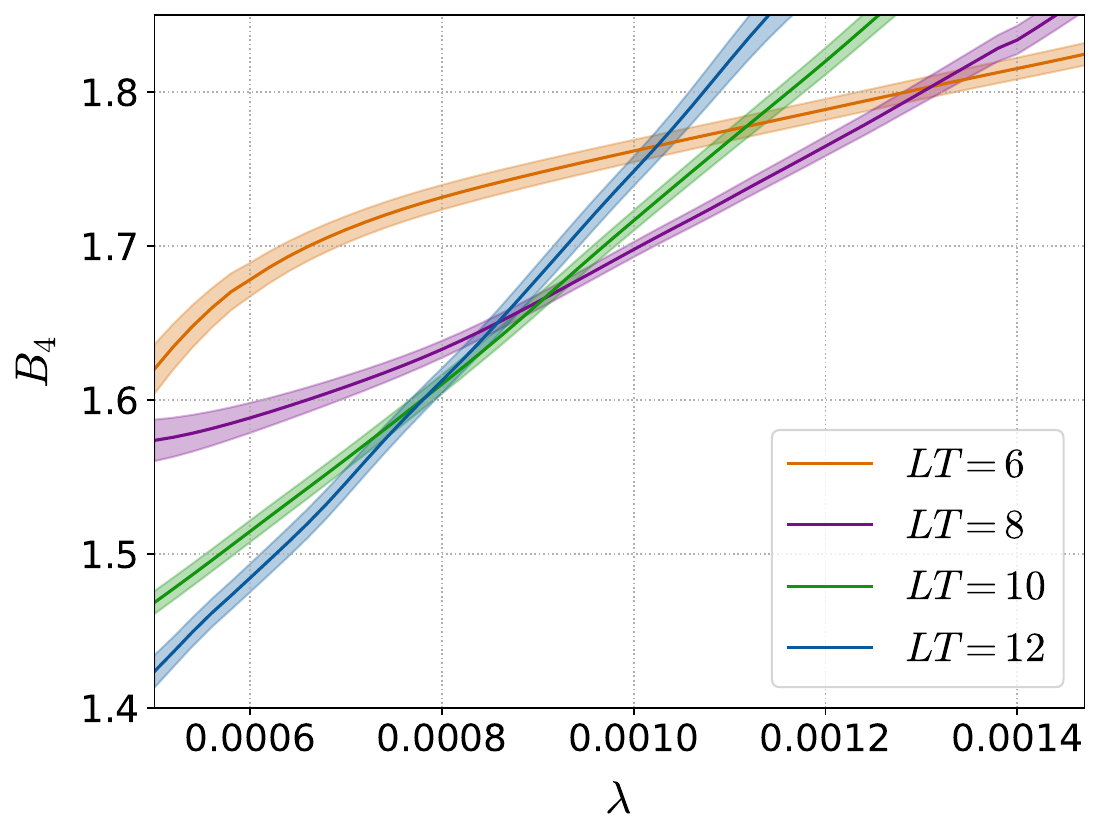}
 \caption{
   $\lambda$ dependence of the Binder cumulant near the CP in the heavy-quark region at $N_t=8$~\cite{prep}.
 }
 \label{fig:8}
\end{figure}

In Fig.~\ref{fig:8}, we finally show our preliminary result of $B_4$ at $N_t=8$ at LO~\cite{prep}. We note that this numerical result does not include the NLO terms. From the figure, one finds that the crossing point of $B_4$ converges to $b_4$ in Eq.~\eqref{eq:b4nu} as $LT$ becomes larger.

\section{Conclusions}

In this proceeding, we performed the Binder-cumulant analysis of the CP in the heavy-quark QCD~\cite{Kiyohara:2021smr,prep}. In our numerical analysis, we employed the HPE to reduce the numerical costs, where the LO terms are included in the Monte-Carlo simulations and the effects of the NLO terms are incorporated by reweighting.
We found that this method is quite effective in reducing statistical
errors by avoiding the overlapping problem of the reweighting method.

From the Binder cumulant analysis, we found that the crossing point $b_4$ and the critical exponent $\nu$ tend to converge to the values in the $Z(2)$ 
and the value of the Polyakov-loop Binder cumulant $B_4$ at the critical point is consistent with the
$Z(2)$ universality class when $LT\ge9$ data are used for the analysis.
On the other hand, a statistically significant deviation from
the $Z(2)$ scaling is observed when the data at
$LT=8$ is included, which suggests that this spatial volume
is not large enough to apply the finite-size scaling.

\vspace{5mm}
\noindent\textbf{Acknowledgments}

This work was supported by in part JSPS KAKENHI (Grant Nos.~JP19H05598, JP20H01903, JP21K03550, JP22K03593, JP22K03619), HPCI System Research project (Project ID: hp200013, hp200089, hp210012, hp210039, hp220020, hp220024), Joint Usage/Research Center for Interdisciplinary Large-scale Information Infrastructures in Japan (JHPCN) (Project ID: jh200010, jh200049), and the Research proposal-based use at the Cybermedia Center, Osaka University, as well as the Multidisciplinary Cooperative Research Program of the Center for Computational Sciences, University of Tsukuba.

\end{document}